\documentclass{emulateapj}
\usepackage{apjfonts}
\slugcomment{Draft Version}

\begin{document}


\title{Innermost collimation structure of the M87 jet down to $\sim$ten
Schwarzschild radii}

\author{Kazuhiro Hada\altaffilmark{1,2,3}, Motoki Kino\altaffilmark{4}, Akihiro
Doi\altaffilmark{4}, Hiroshi Nagai\altaffilmark{2}, Mareki
Honma\altaffilmark{2,3}, Yoshiaki Hagiwara\altaffilmark{2,3}, \\ Marcello
Giroletti\altaffilmark{1}, Gabriele Giovannini\altaffilmark{1,5} and Noriyuki
Kawaguchi\altaffilmark{2,3}}

\affil{$^1$INAF Istituto di Radioastronomia, via Gobetti 101, I-40129
Bologna, Italy}

\affil{$^2$National Astronomical Observatory of Japan,
Osawa, Mitaka, Tokyo 181-8588, Japan} 

\affil{$^3$Department of Astronomical Science, The Graduate University
for Advanced Studies (SOKENDAI), 2-21-1 Osawa, Mitaka, Tokyo 181-8588,
Japan}

\affil{$^4$Institute of Space and
Astronautical Science, Japan Aerospace Exploration Agency, 3-1-1
Yoshinodai, Chuo, Sagamihara 252-5210, Japan} 

\affil{$^5$Dipartimento di Fisica e Astronomia, Universit\`a di Bologna, via
Ranzani 1, I-40127 Bologna, Italy}

\begin{abstract}
We investigated the detailed inner jet structure of M87 using the Very Long
Baseline Array data at 2, 5, 8.4, 15, 23.8, 43, and 86~GHz, especially focusing on
the multi-frequency properties of the radio core at the jet base. First, we
measured a size of the core region transverse to the jet axis, defined as $W_{\rm
c}$, at each frequency $\nu$, and found a relation between $W_{\rm c}$ and $\nu$
as $W_{\rm c}(\nu) \propto \nu^{-0.71\pm0.05}$. Then, by combining $W_{\rm
c}(\nu)$ and the frequency dependence of the core position $r_{\rm c}(\nu)$, which
was obtained by our previous study, we have constructed a collimation profile of
the innermost jet $W_{\rm c}(r)$ down to $\sim$10 Schwarzschild radii ($R_{\rm
s}$) from the central black hole. We found that $W_{\rm c}(r)$ smoothly connects
with the width profile of the outer edge-brightened, parabolic jet, and then
follows a similar radial dependence down to several tens of $R_{\rm s}$. Closer to
the black hole, the measured radial profile suggests a possible change of the jet
collimation shape from the outer parabolic one, in which the jet shape tends to
become more radially-oriented. This could be related to a magnetic collimation
process or/and interections with surrounding materials at the jet base. The
present results shed light on the importance of higher-sensitivity/resolution
imaging studies for M87 at 86, 43 and also 22~GHz, and should be examined more
rigorously.
 
\end{abstract}


\section{Introduction} 
Formation of relativistic jets in active galactic nuclei (AGN) is one of the
biggest challenges in astrophysics. The radio galaxy M87 is one of the closest
examples of this phenomena, and its jet has been investigated across the entire
electromagnetic spectrum over years~\citep[e.g.,][]{owen1989, biretta1999,
harris2006, abramowski2012}. Due to its proximty
\citep[$D=16.7$~Mpc;][]{jordan2005} and a large estimated mass of the central
black hole~\citep[$M_{\rm BH} \simeq (3-6) \times
10^9~M_{\odot}$;][]{macchetto1997, gebhardt2009, walsh2013}\footnote{In this
Letter we adopt $M_{\rm BH} = 6.0\times 10^9M_{\odot}$ along with \citet{hada2011,
hada2012}.}, 1 milliarcsecond attains 0.08~pc or 140~Schwarzschild radii $(R_{\rm
s})$, providing an unprecedented opportunity to probe the jet formation processes
at its base.

The inner jet structure of M87 has been intensively investigated with
Very-Long-Baseline-Interferometry (VLBI). \citet{junor1999} discovered a
broadening of the jet opening angle at $\sim$100~$R_{\rm s}$ from the radio core
with an edge-brightened structure, and this was later confirmed in several
works~\citep{ly2004, dodson2006, krichbaum2006, ly2007, kovalev2007}.  More
recently, \citet[][hereafter AN12]{asada2012} discovered a maintanance of a
parabolic structure between a few $100~R_{\rm s}$ and $10^5~R_{\rm s}$ from the
core with a transition into a conical streamline above a location of
$\sim10^6~R_{\rm s}$. These results provide compelling evidence that the
collimation region is beginning to be resolved near the base of this jet.

However, the jet structure within $\sim$100$~R_{\rm s}$ remains unclear. Probing
the innermost region is essential to directly test theoretical models of
relativistic jet formation. Because the radio core at the jet base corresponds to
an optically-thick surface of synchrotron emission~\citep{bk1979}, previous
studies were not possible to determine the location of the central engine
relative to the radio core, preventing them from estimating the exact collimation
profile in this region.

Recently, we have overcome this difficulty~\citep[][hereafter H11]{hada2011}. By
measuring the core position shift~\citep{konigl1981, lobanov1998} with
multi-frequency phase-referencing Very-Long-Baseline-Array (VLBA) observations, we
have constrained the location of the central engine of the M87 jet as
$\sim$20$~R_{\rm s}$ upstream of the 43-GHz core. This allows us to probe the
radial profile of the jet structure as a function of the distance from the central
engine. Moreover, the determination of the frequency dependence of the core
position $r_{\rm c}\propto \nu^{-0.94\pm0.09}$ enables us to reveal the
structure of the innermost jet by investigating the multi-frequency properties of
the core. Indeed, the recent VLBI observation of M87 at 230~GHz detected a compact
core which is comparable to the size of event horizon~\citep{doeleman2012}, being
consistent with the core of this jet to be located in the black hole vicinity.

In this paper, we explore the collimation profile of the inner jet of M87,
especially based on the multi-frequency properties of the VLBI core as well as the
edge-brightened jet. The data analysis is described in the next section. In
section 3, we show our new results. In the final section, we discuss the derived
inner jet structure.

\begin{table*}[htbp]
 \begin{minipage}[t]{1.0\textwidth}
 \centering
  \caption{Multi-frequency VLBA observations of M87}
  \scalebox{1.00}{
  \begin{tabular}{lllccccccc}
   \hline
   \hline
   Freq. & Date & Beam $\theta_{\rm bm}$ & $I_{\rm rms}$ & \multicolumn{ 6}{c}{JMFIT Gaussian}       \\\cline{5-10}
   band  & & size, P.A. &  &  & $\theta_{\rm maj}$ & $\theta_{\rm min}$ & P.A. &
   $\theta_{20^{\circ}} (\equiv W_{\rm c})$ & $\frac{\theta_{20^{\circ}}}{\theta_{\rm bm, 20^{\circ}}}$ \\
   (GHz) &             & (mas)$^2$, (deg.)       & $\left(\frac{\rm mJy}{\rm beam}\right)$ &     & (mas) & (mas) & (deg.) & (mas) &  \\
   &             &  (a)                      & (b)  &      & (c)      & (d)   & (e)         & (f)        & (g) \\
   \hline
   2.3 & 2010 Apr  8 & 6.01 $\times$ 2.92, $-3$  & 0.79 &      & $3.26\pm 0.08$ & $< 1.06$ & $286\pm 3$ &  $<1.06$ & $<0.20$ \\
       & 2010 Apr 18 & 5.97 $\times$ 3.00, $-4$  & 0.66 &      & $3.47\pm 0.08$ & $< 1.06$ & $287\pm 3$ &  $<1.06$ & $<0.20$ \\\cline{5-10}
       &             &                           &      & Ave. & 3.37 & $< 1.06$ & 287 & $<1.06$ & $<0.20$ \\
       &             &                           &      &      & & & & & \\
   5.0 & 2010 Apr 8  & 2.63 $\times$ 1.36, 1     & 0.43 &      & $1.36\pm 0.06$ & $0.57\pm 0.16$ & $285\pm 4$ & $0.58\pm 0.17$ & 0.25\\
       & 2010 Apr 18 & 2.70 $\times$ 1.31, 0     & 0.37 &      & $1.38\pm 0.06$ & $0.56\pm 0.13$ & $288\pm 3$ & $0.56\pm 0.14$ & 0.24 \\\cline{5-10}
       &             &                           &      & Ave. & 1.37 & 0.57 & 287 & 0.57 & 0.25 \\
       &             &                           &      &      & & & & &  \\
   8.4 & 2009 May 23 & 1.48 $\times$ 0.55, $-9$  & 0.21 &      & $0.61\pm 0.07$ & $0.38\pm 0.05$ & $292\pm 5$ & $0.38\pm 0.06$ & 0.40 \\
       & 2010 Apr  8 & 1.67 $\times$ 0.84, 2     & 0.37 &      & $0.84\pm 0.04$ & $0.42\pm 0.06$ & $295\pm 4$ & $0.42\pm 0.07$ & 0.29 \\
       & 2010 Apr 18 & 1.63 $\times$ 0.85, $-3$  & 0.34 &      & $0.87\pm 0.05$ & $0.40\pm 0.08$ & $295\pm 4$ & $0.40\pm 0.09$ & 0.29 \\\cline{5-10}
       &             &                           &      & Ave. & 0.77 & 0.40 & 294 & 0.40 & 0.33 \\
       &             &                           &      &      & & & & & \\
   15  & 2009 Jan 7  & 1.04 $\times$ 0.47, $-3$  & 0.82 &      & $0.45\pm 0.02$ & $0.28\pm 0.01$ & $298\pm 5$ & $0.28\pm 0.02$ & 0.34 \\
       & 2009 Jul 5  & 1.00 $\times$ 0.46, $-8$  & 0.63 &      & $0.39\pm 0.05$ & $0.26\pm 0.03$ & $305\pm 7$ & $0.26\pm 0.06$ & 0.35 \\
       & 2010 Feb 11 & 1.00 $\times$ 0.43, $-2$  & 1.07 &      & $0.41\pm 0.02$ & $0.21\pm 0.04$ & $286\pm 2$ & $0.21\pm 0.04$ & 0.27 \\
       & 2010 Apr 8  & 0.94 $\times$ 0.48, $-$5  & 0.59 &      & $0.47\pm 0.02$ & $0.25\pm 0.04$ & $300\pm 5$ & $0.25\pm 0.04$ & 0.33 \\
       & 2010 Apr 18 & 0.92 $\times$ 0.49, $-10$ & 0.44 &      & $0.49\pm 0.02$ & $0.24\pm 0.02$ & $298\pm 3$ & $0.24\pm 0.03$ & 0.33 \\
       & 2010 Sep 29 & 1.00 $\times$ 0.45, $-2$  & 0.72 &      & $0.42\pm 0.07$ & $0.26\pm 0.03$ & $307\pm 4$ & $0.27\pm 0.08$ & 0.35 \\
       & 2011 May 21 & 0.93 $\times$ 0.43, $-5$  & 0.74 &      & $0.51\pm 0.05$ & $0.28\pm 0.05$ & $302\pm 5$ & $0.29\pm 0.07$ & 0.38 \\\cline{5-10}
       &             &                           &      & Ave. & 0.45 & 0.25 & 299 & 0.26 & 0.34 \\
       &             &                           &      &      & & & & & \\
  23.8 & 2010 Jan 18 & 0.60 $\times$ 0.29, $-8$  & 0.76 &      & $0.25\pm 0.04$ & $0.20\pm 0.02$ & $300\pm 3$ & $0.20\pm 0.04$ & 0.42 \\
       & 2010 Apr 4  & 0.63 $\times$ 0.29, $-4$  & 0.68 &      & $0.27\pm 0.02$ & $0.18\pm 0.03$ & $313\pm 7$ & $0.19\pm 0.04$ & 0.38 \\
       & 2010 Apr 8  & 0.54 $\times$ 0.29, $-5$  & 1.36 &      & $0.24\pm 0.03$ & $0.18\pm 0.04$ & $326\pm 6$ & $0.20\pm 0.05$ & 0.43 \\
       & 2010 Apr 18 & 0.57 $\times$ 0.28, $-13$ & 1.12 &      & $0.25\pm 0.02$ & $0.20\pm 0.03$ & $327\pm 3$ & $0.21\pm 0.04$ & 0.54 \\
       & 2010 May 1  & 0.62 $\times$ 0.28, $-9$  & 0.73 &      & $0.29\pm 0.02$ & $0.19\pm 0.03$ & $315\pm 6$ & $0.20\pm 0.04$ & 0.42 \\
       & 2010 May 15 & 0.64 $\times$ 0.29, $-12$ & 0.80 &      & $0.31\pm 0.03$ & $0.20\pm 0.04$ & $317\pm 4$ & $0.21\pm 0.05$ & 0.48 \\
       & 2010 May 30 & 0.62 $\times$ 0.28, $-11$ & 0.83 &      & $0.28\pm 0.02$ & $0.16\pm 0.05$ & $314\pm 5$ & $0.17\pm 0.05$ & 0.38 \\\cline{5-10}
       &             &                           &      & Ave. & 0.27 & 0.19 & 316 & 0.20 & 0.44 \\
       &             &                           &      &      & & & & &   \\
   43  & 2009 Mar 13 & 0.29 $\times$ 0.13, $-6$  & 0.55 &      & $0.15\pm 0.01$ & $0.13\pm 0.02$ & $313\pm 9$ & $0.13\pm 0.02$ & 0.61 \\
       & 2010 Jan 18 & 0.29 $\times$ 0.13, 2     & 1.01 &      & $0.11\pm 0.02$ & $0.10\pm 0.02$ & $262\pm 3$ & $0.10\pm 0.03$ & 0.39 \\
       & 2010 Apr 8  & 0.26 $\times$ 0.13, $-3$  & 0.91 &      & $0.13\pm 0.02$ & $0.11\pm 0.02$ & $354\pm 4$ & $0.13\pm 0.03$ & 0.57 \\
       & 2010 Apr 18 & 0.26 $\times$ 0.13, $-8$  & 1.12 &      & $0.13\pm 0.01$ & $0.11\pm 0.01$ & $358\pm 3$ & $0.13\pm 0.01$ & 0.63 \\
       & 2010 May 1  & 0.29 $\times$ 0.14, $-6$  & 0.89 &      & $0.11\pm 0.02$ & $0.11\pm 0.01$ & $318\pm 19$& $0.11\pm 0.02$ & 0.49 \\
       & 2010 May 15 & 0.29 $\times$ 0.14, $-9$  & 0.95 &      & $0.12\pm 0.01$ & $0.10\pm 0.02$ & $254\pm 6$ & $0.11\pm 0.02$ & 0.48 \\
       & 2010 May 30 & 0.30 $\times$ 0.15, $-6$  & 1.07 &      & $0.13\pm 0.01$ & $0.09\pm 0.01$ & $257\pm 8$ & $0.10\pm 0.02$ & 0.42 \\\cline{5-10}
       &             &                           &      & Ave. & 0.13 & 0.11 & 280 & 0.11 & 0.51 \\
       &             &                           &      &      & & & & &  \\
   86  & 2007 Feb 18 & 0.25 $\times$ 0.08, $-18$ & 6.47 &      & $0.079\pm 0.021$ & $0.065\pm 0.023$ & $52\pm 23$ & $0.074\pm 0.032$ & 0.64 \\
   \hline
  \end{tabular}
  }
 \medskip
 \end{minipage}
 \label{tab:tab1} Note: (a) uniformly-weighted beam size; (b) image noise level;
 (c),(d),(e) FWHM sizes of major/minor axes and position angle of derived model;
 (f) projected FWHM size of the model in $\rm{P.A.}=20^{\circ}$; (g) ratio of the
 Gaussian size devided by the beam size in $\rm{P.A.}=20^{\circ}$.
\end{table*}

   \begin{figure}[htbp]
   \centering
   \includegraphics[angle=0,width=1.0\columnwidth]{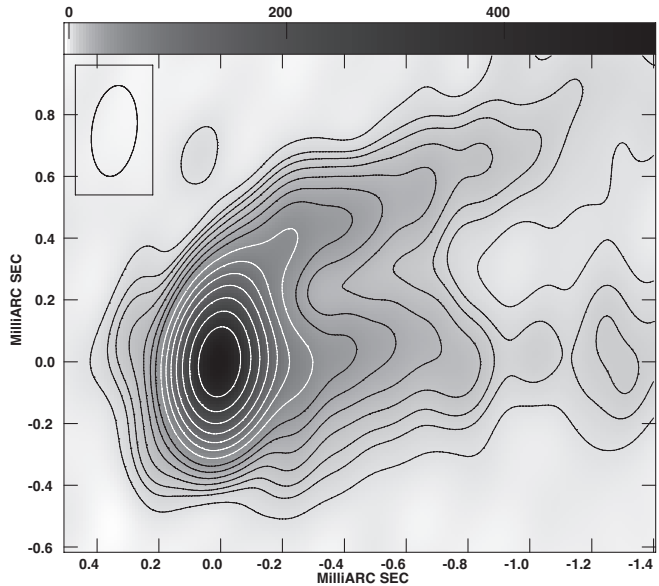}
   \caption{Uniformly-weighted, averaged VLBA image of M87 at 43~GHz. Contours
   start from 3$\sigma$ image rms level and increasing by factors of 1.4.}
   \label{fig:image}
   \end{figure}

\section{Observations and Data Reduction}
We observed M87 with VLBA at 2, 5, 8.4, 15, 23.8 and 43~GHz on 2010 April 8 and
18. These are the same data presented in H11, where we investigated the core shift
of M87 using the phase-referencing technique relative to the nearby radio source
M84. Details of the observations and the data reductions processes are described
in H11.

To better constrain the averaged multi-frequency properties of the core and the
inner jet, we also analyzed VLBA archival data at 8.4, 15, 23.8 and 43~GHz. We
selected data observed after 2009 with sufficiently good qualities (all 10
stations participated, good $uv$-coverages, and thus high angular resolutions
obtained). Initial data calibration before imaging was performed using the
National Radio Astronomy Observatory (NRAO) Astronomical Image Processing System
(AIPS) based on the standard VLBI data reduction procedures. These data were not
acquired with phase-referencing mode.

Moreover, we added one VLBA archival data set at 86~GHz, which allows us to probe
the inner jet even closer to the black hole because of its higher transparency and
resolution. While several VLBA observations of M87 have been performed at 86~GHz,
we selected the data observed in 2007 February, because this is the only 86~GHz
data at present for which a reliaiable self-calibrated image has been published
with VLBA alone (peak-to-noise ratio of the self-calibrated image higher than
$\sim$70), as shown in \citet{rioja2011}. The observation is operated at 8
stations without Saint-Croix and Hancock. We analyzed this data based on the
procedures described in their paper.

Images were created in DIFMAP software with iterative phase/amplitude
self-calibration. We used uniform weighting scheme to produce high resolution
images.

\section{Results}
The M87 jet was clearly detected for all of the analyzed data. In Figure~1 we show
a representative image of M87 at 43~GHz, which was made by stacking the seven sets
of data. We confirmed the jet to be characterized by the compact core with the
edge-brightened structure directing an overall position angle of
P.A.$\sim$290$^{\circ}$.

\subsection{Model fitting on the core region}

In the present study, we aim at measuring the width of the innermost jet. For this
purpose, we fitted a single elliptical Gaussian model to each image with the AIPS
task JMFIT, and derived deconvolved parameters of the core region. Note that the
derived Gaussian size in this simple modeling could yield a larger size than that
of the true core (i.e., optical depth $\sim1$ surface) in the jet propagation
direction, because of blending of the optically-thin jet part. Also in the
transverse direction to the jet axis, this method would give a total size of the
true core plus surrounding emission, if the core region has a sub-structure that
is not resolved by VLBA baselines in this direction~\citep{dodson2006}. However,
here we are interested in measuring such a entire width of the innermost jet.

The results are summarized in Table~\ref{tab:tab1}. Most of the derived values
(especially for minor axes) are smaller than the beam size for each
data. Nonetheless, it is known that such sizes are measurable when the emitting
region is sufficeintly bright and robust self-calibration using as many as 10 VLBA
stations (8 at 86~GHz) can calibrate the fringe visibilities accurately. For the
M87 core, the derived sizes along the minor axes at 5, 8.4, 15, 23.8, 43 and
86~GHz correspond to amplitude decreases of 15\%, 18\%, 25\%, 30\%, 33\% and 30\%
at $\sim$60\% of the longest baseline (80M$\lambda$, 140M$\lambda$, 240M$\lambda$,
380M$\lambda$, 700M$\lambda$ and 1100M$\lambda$ respectively, which yield
effective angular resolutions in ${\rm P.A.=20^{\circ}}$). These decreases are
sufficiently larger than a typical VLBA amplitude calibration accuracy of
$\sim$5\% \citep[analogous discussion is presented in][]{kellermann1998}. At
86~GHz, previous Global-Millimeter-VLBI-Array observations report similar sizes to
the present value in Table~\ref{tab:tab1} \citep[$\lesssim 50$~$\mu$as or
$99\pm21$~$\mu$as;][respectively]{krichbaum2006, lee2008}.

We estimated the standard errors of the derived Gaussian parameters for each data
as follows. Formally, statistical size uncertainties purely based on
SNR~\citep[size of the fitted model devided by its peak-to-noise ratio;
e.g.,][]{fomalont1999} results in very small values for M87 (a level of a few
$\mu$as or smaller) because the core is bright at each frequency (peak-to-noise
$>70\sim1000$). However in practice, the model parameters are more strongly
affected by imperfect CLEAN/deconvolution processes under limited
$uv$-coverages. Then, we devided each individual data set into three subsets with
equal integration time, and repeated deconvolution and JMFIT processes for each
data individually. Through this procedure, we obtained three sets of
quasi-independent fitting results for each epoch, and the rms scatters can be
calculated for each parameter (i.e., major/minor axes and P.A.). These scatters
are adopted as realistic errors for each model parameter in Table~\ref{tab:tab1}.

As an additional check, we conducted a fake visibility analysis to examine how
precisely sizes smaller than beam sizes can be recovered~\citep[a similar analysis
is described in][]{taylor2004}; using the AIPS task UVCON, we created fake
visibilities which are equivalent to the derived Gaussian parameters at each
frequency with similar $uv$-coverages of the actual observations. We produced 10
independent visibility data sets at each frequency by adding random noises at a
level seen in the actual observations, repeated CLEAN/deconvolution and JMFIT on
each data, and calculated rms scatters of the recovered model parameters at each
frequency. We confirmed that these scatters were smaller than one-third of the
quoted errors in Table~\ref{tab:tab1} at frequencies smaller than 43~GHz, and less
than one-half at 86~GHz.

We note that, only regarding the minor axis of the 2-GHz core, JMFIT derived a
quite small size (less than 15\% of the beam size), which indicates a marginal
constraint.  So in Table~\ref{tab:tab1} we instead set one-fifth of the beam size
as an upper limit, which corresponds to $\sim$5\% amplitude decrease at the
longest baseline in this direction. 

   \begin{figure}[bbb]
   \centering \includegraphics[angle=0,width=1.0\columnwidth]{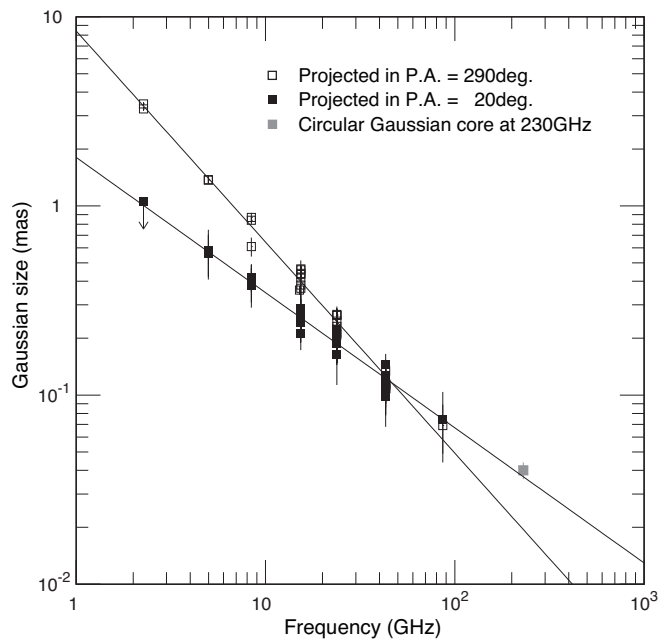}
   \caption{Gaussian sizes for the core region as a function of frequency. Sets of
   filled/open rectangle indicates projected sizes in ${\rm
   P.A.}=20^{\circ}/290^{\circ}$. Two lines are power-law fit to each distribution
   using 8.4, 15, 23.8 and 43~GHz data. A gray rectangle indicates the derived
   core size at 230~GHz~\citep{doeleman2012}. } \label{fig:coresize}
   \end{figure}

   \begin{figure*}[htbp]
   \centering \includegraphics[angle=0,width=0.75\textwidth]{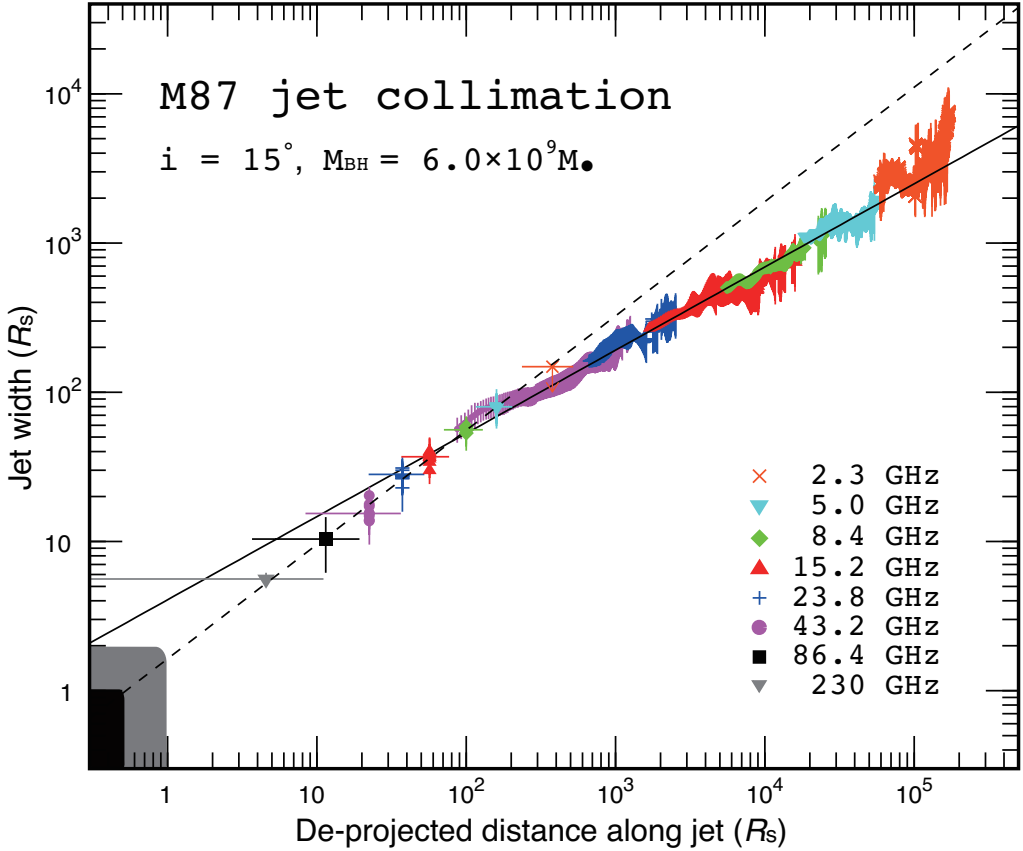}
   \caption{Jet width profile of M87 as a function of distance. A jet viewing
   angle of $i=15^{\circ}$ and a black hole mass of $M_{\rm
   BH}=6.0\times10^9M_{\odot}$ are adopted. Densely-sampled part indicates the
   width profile of the edge-brightened jet measured at multiple frequency with
   errors. Discretely-sampled part indicates the FWHM size profile for the core
   region projected in P.A=20$^{\circ}$ ($W_{\rm c}$). The core positions and the
   errors are estimated based on the previous astrometry
   result~\citep{hada2011}. A solid line indicates the best fit solution for the
   edge-brightened jet width $W_{\rm j}\propto r^{0.56}$, while a dashed line
   indicates $r^{0.76}$. The grey/black rectangles at the bottom left corner of
   the figure represent surfaces (event horizons) for
   non-spinning/maximally-spinning black holes, respectively.}
   \label{fig:jetwidth}
   \end{figure*}

In Figure~\ref{fig:coresize}, we show a Gaussian size projected along ${\rm
P.A.=20^{\circ}}$ as a function of frequency (hereafter we denote $W_{\rm
c}$)\footnote{Here $W_{\rm c}$ is defined as a FWHM of the derived ellipical
Gaussians when sliced along a position angle of $20^{\circ}$ i.e., $W_{\rm c}
\equiv \left[\frac{\theta_{\rm maj}^2 \theta_{\rm min}^2 (1 + \tan^2 (20^{\circ} -
{\rm P.A.})}{\theta_{\rm min}^2 + \theta_{\rm maj}^2 \tan^2 (20^{\circ} - {\rm
P.A.})}\right]^{1/2}$, where $\theta_{\rm maj}, \theta_{\rm min}$ and
P.A. indicate values in columns (c), (d) and (e) of Table~1. The errors in this
direction, shown in column (g) of Table~\ref{tab:tab1}, are calculated from usual
error propagation analysis using the errors $\theta_{\rm maj}, \theta_{\rm min}$
and P.A..}. This projection is perpendicular to an overall jet axis, correnponding
to a direction of the jet width. We found $W_{\rm c}$ to be clearly frequency
dependent, becoming smaller as frequency increases. To determine an averaged
frequency dependence of $W_{\rm c}$, we fitted a power law function to this plot
using the data at 5, 8.4, 15, 23.8, 43 and 86~GHz (2~GHz data are excluded because
of upper limits). We found the best-fit solution to be $W_{\rm c}(\nu) \propto
\nu^{\xi}$ where $\xi = -0.71\pm0.05$. Interestingly, when this frequency dependence
is extended toward higher frequency, its extrapolated size appears to result in a
similar value to the measured size by the recent 230~GHz VLBI
experiment~\citep{doeleman2012}, which is determined with a circular Gaussian
fit. For reference, we also show a Gaussian size distribution projected along a
jet propagation direction, but this does not seem to fit the 230~GHz core size.

\subsection{Jet width measurements}

In Figure~\ref{fig:jetwidth}, we show the radial profile of the M87 jet width as a
function of (de-projected) distance along jet \citep[assuming a jet inclination
angle $i=15^{\circ}$;][]{biretta1999, perlman2011}. Here we investigated the jet
width profile in the following two procedures: (1) measurements of $W_{\rm j}(r)$;
for the region where the jet is clearly resolved into a two-humped shape at each
frequency, we made transverse slices of the jet every 0.01$\sim$0.5~mas distance
along the jet, and each of the two-humped structure was fitted by a
double-Gaussian function. We then defined the separation between the outer sides
of the half-maximum points of the two Gaussains as jet width. When the jet at a
frequency becomes single-humped toward the upstream region, we measured the jet
width using higher frequency images because the jet is clearly resolved again into
a two-humped shape at higher resolutions. By using the images between 2 and
43~GHz, such a measurement was repeated over a distance of $\sim$10$^{5}$~$R_{\rm
s}$ down to $\sim$100~$R_{\rm s}$ along the jet. This process is basically the
same as that used in AN12, but we exclude measurements for the single-humped
region. We aligned radial positions of $W_{\rm j}(r)$ profiles between different
frequencies by adding the amount of core shift measured in H11 (described
below). This amount and the associated position error at each frequency provide
only tiny fractions of the distance where $W_{\rm j}(r)$ was measured at each
frequency (a level of $10^{-1\sim-2}$ at 43~GHz to $10^{-3}$ at 2~GHz), so
horizontal error bars for $W_{\rm j}(r)$ are removed in
Figure~\ref{fig:jetwidth}. At 86~GHz, we could not perform reliable measurements
of the jet width because the edge-brightened jet was only marginally imaged at a
level of (2--3)$\sigma$. (2) measurements of $W_{\rm c}(r)$; closer to the central
engine, we further constructed a radial distribution of $W_{\rm c}$. Because our
previous astrometric study H11 measured locations of the cores as 41, 69, 109,
188, 295 and 674~$\mu$as at 43, 23.8, 15, 8.4, 5 and 2~GHz from the convergence
point of the core shift (in R.A. direction or P.A$=270^{\circ}$), we can set their
de-projected distances along the jet (P.A.$=290^{\circ}$) as 24, 40, 63, 108, 170
and 388~$R_{\rm s}$ for $i=15^{\circ}$, respectively. For 86/230~GHz cores, we can
also set their de-projected positions as 12 and 5~$R_{\rm s}$ by assuming the same
asymptotic relation ($r_{\rm c}\propto\nu^{-0.94}$; H11) for the upstream of the
43~GHz core. Here we assume that the central engine is located at the convergence
point of the core shift specified in H11.

We confirmed that the edge-brightened region is well-expressed as a parabolic
structure of $W_{\rm j}(r)\propto r^{0.56\pm0.03}$ (solid line in Figure~3). This
is in good agreement with their finding $r^{0.58\pm0.02}$ in AN12. For the region
around $\sim$100~$R_{\rm s}$, where the independent measurements of $W_{\rm j}$
and $W_{\rm c}$ are overlapped each other, $W_{\rm c}$ at 5 and 8,4~GHz smoothly
connect with $W_{\rm j}$ at 43~GHz. The combination of the core size ($W_{\rm
c}\propto \nu^{\xi}$ where $\xi=-0.71\pm0.05$) and core position ($r\propto
\nu^{\alpha}$ where $\alpha={-0.94\pm0.09}$; H11) yields a radial dependence of
$W_{\rm c}$ as $W_{\rm c}(r) \propto r^{\frac{\xi}{\alpha}} = r^{0.76\pm0.13}$
(dotted line in Figure~3), which is slightly steeper than that of the outer jet
$W_{\rm j}(r)$, although the uncertainty is still large. In the present result, it
is still difficult to distinguish whether $W_{\rm c}$ at 5, 8.4, 15 and 23.8~GHz
are on the solid line or the dashed line due to their position uncertainties in
addition to those of sizes. On the other hand, $W_{\rm c}$ at 43 and 86~GHz tends
to be below the solid line. At 230~GHz, the exact profile cannot be discriminated
again because the data is totally dominated by its position uncertainty.

We note that the two methods used for $W_{\rm j}(r)$ (a double-Gaussain fit on
each slice image) and $W_{\rm c}(r)$ (based on a single two-dimensional elliptical
Gaussian model on each 2-D image) are different from each other. Nevertheless, the
observed consistencies of the values between the two methods were confirmed in the
overlapped region, indicating that $W_{\rm c}$ is actually a good tracer for the
width of the innermost jet region.

\section{Discussion}
Probing the collimation profile of the M87 jet is crucial to understand the
formation processes of relativistic jets.  Although the actual energetics of the
M87 jet is still under debate~\citep[e.g.,][]{abdo2009}, here we focus on the
framework of magnetohydrodynamic (MHD) jets, because it is widely explored as the
most successful scenario for jet production.

\subsection{$r\gtrsim100~R_{\rm s}$ Region}
Theoretically, the shape of a magnetized jet is determined by the detailed force
balance across the poloidal field lines. This is described as the trans-field
equation, which was originally derived by \citet{okamoto1975} for steady,
axisymmetric, rotating magnetized flow in a gravitational field without external
pressure. Later, it is invoked that magnetic hoop stresses associated with
toroidal field lines play a major role to realize a global collimation of a
magnetized jet~\citep[]{bp1982, heyvaerts1989, sakurai1985, chiueh1991}.

Here we observationally confirmed that the M87 jet is well characterized as a
parabolic collimation profile between $\sim$100 and $\sim$10$^{5}~R_{\rm s}$. This
is consistent with the prior work by AN12, where they also found a transition into
a conical shape above $\sim$$10^{6}~R_{\rm s}$. Regarding formation of a parabolic
shape, recent theoretical studies indicate the importance of external pressure
support at the jet boundary. \citet{okamoto1999} analytically showed that hoop
stresses alone would not realize global collimation, and numerical studies also
come along this line~\citep[e.g.,][]{nakamura2006, komissarov2007,
komissarov2009, toma2013}.
 \citet{komissarov2009} shows that when the external gas
pressure follows $p_{\rm ext}\propto r^{-a}$ where $a \lesssim 2$, the jet
maintains parabolic as $W_{\rm j}\propto r^{a/4}$, whereas for $a > 2$ the jet
eventually becomes conical due to insufficient external support. If the observed
radio emission of M87 traces the exterior edge of a magnetized jet, the measured
width profile suggests $a \sim2$. As a source of such confinement medium, AN12
propose an intersteller medium bounded by the gravitational influence of the
central black hole, such as a giant ADAF~\citep{narayan2011}. We add to note that
a purely hydrodynamic (HD) jet is also possible to produce the gradual parabolic
collimation of M87~\citep[e.g.,][]{bromberg2009}.

Interestingly, at the end of the collimation region, both AN12 and
\citet{bromberg2009} suggest HST-1 (a peculiar knot at a deprojected distance
$\gtrsim$120~pc or $2\times10^5R_{\rm s}$) as a stationary shocked feature
resulting from overcollimation of the inner jet, which is originally proposed by
\citet{stawarz2006} and \citet{cheung2007} to explain the broadband flaring
activity. While our recent kimematic observations of HST-1 shows clear evidence of
superluminal motions of the overall structure, a weak, slowly-moving feature is
also found in the upstream of HST-1~\citep{giroletti2012}, which indeed could be
related to a recollimation shock.

  \begin{figure}[ttt]
   \centering \includegraphics[angle=0,width=1.01\columnwidth]{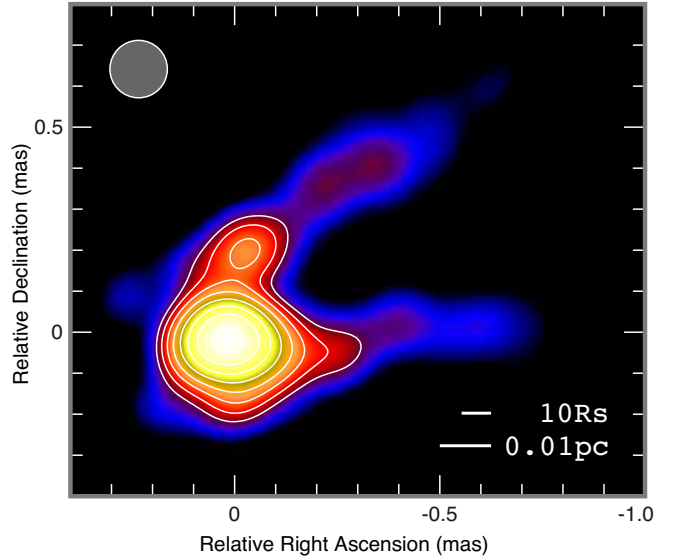}
   \caption{The same image as Fig.~1 but convolved with a circular Gaussian beam
   of 0.14~mas diameter, yielding roughly twice higher resolution in north-south
   direction. Contours in the inner region starts from 20$\sigma$ image rms and
   increasing by factors of 1.4. The length bars correspond to projected scale.}
   \label{fig:m87q_sr}
  \end{figure}

\subsection{$r\lesssim100~R_{\rm s}$ Region}

For the first time, we have revealed a detailed collimation profile down to $r\sim
10~R_{\rm s}$ by investigating the multi-frequency properties of the radio
core. An intriguing point here is that the measured collimation profile suggests a
possible tendency of a wider jet opening angle around $\sim$10 to $\sim$100~$R_{\rm
s}$ from the central engine.

Since the two methods of our width measurements are switched around
$r\sim100~R_{\rm s}$, one could speculate that the profile change is related to
some systematic effects due to the different methods. Then, to check the
two-humped jet shape further close to the core ($r\lesssim 100~R_{\rm s}$) more
clearly, we created a 43-GHz image convolved with a slightly higher resolution (a
circular beam of 0.14~mas diameter). The image is shown in Figure 4. Above
$\sim$0.4~mas downstream of the core (de-projected distance $\sim$220~$R_{\rm s}$
for $i=15^{\circ}$), where the jet is parabolic on logarithmic distance, the two
ridges are already oriented into a similar direction (dark-blue region in Figure
4). On the other hand, the opening angle made by the two ridges appear to broaden
more rapidly within $\sim$0.3~mas of the core (the region with contours in Figure
4), resulting in a more radially-oriented structure near the base. Such a tendency
is consistent with the observed possible transition from the solid line to the
(steeper) dashed line around $r\sim100~R_{\rm s}$ in Figure 3.


A transition of jet collimation profile near the black hole is actually suggested
from some of theoretical aspects. In the framework of relativistic MHD jet models,
most of the energy conversion from magnetic-to-kinetic occurs after the flow
passes through the fast-magnetosonic point~\citep[``magnetic nozzle''
effect;][]{li1992, vlahakis2003}. Beyond this point, the magnetized jet starts to
be collimated asymptotically into a parabolic shape, because the increasing plasma
inertia winds the field lines into toroidal directions and thus amplifies the hoop
stresses~\citep[e.g.,][]{tomimatsu2003}. The radius of the fast point is typically
a few times the light cylinder radius $R_{\rm lc}$~\citep{li1992}, where $R_{\rm
lc}$ is of the order of (1$\sim$5)~$R_{\rm s}$~\citep{komissarov2007}. Thus, if
the M87 jet is magnetically launched, the observed possible transition of the jet
shape around $10\sim100~R_{\rm s}$ could be explained by this process. Moreover,
the jet in this scale is likely to have complicated interactions with surrounding
medium such as accretion flow, corona and disk
wind~\citep[e.g.,][]{mckinney2006}. Their geometries and the local pressure
balance at the jet boundary would affect the initial jet shape. Alternatively,
such a change of the jet shape could happen as an apparent effect due to
projection, if the jet inclination angle is not constant down to the black
hole~\citep{mckinney2013}.

It is interesting to note that a time-averaged dependence $\nu^{-0.71}$ of $W_{\rm
c}$ appears to connect with the 230~GHz core size. However, this apparent
connection should be compared in a cautious manner; the $uv$-coverage used in
\citet{doeleman2012} yields the highest ($\sim$3000M$\lambda$) angular resolution
\textit{along} the jet direction for M87, while $\sim$5 times shorter projected
baselines transverse to the jet. Thus the derived size of $\sim$40~$\mu$as by
their circular Gaussian fit could be more weighted to the structure along the jet,
unless the brightness pattern of the jet base (when projected toward us) is
actually close to a circular shape. To clarify the exact relationship of the core
size at the higher frequency side, the addition of north-south baselines in the
future Event-Horizon-Telescope array is crucial, which can be realized by
including Chilean stations such as ASTE, APEX and ALMA~\citep{broderick2009,
doeleman2009}.



The results presented here have newly shed light on the crucial issues which
should be addressed more rigorously in future observations; where is the exact
location of the profile change between $\sim$10 and $\sim$100~$R_{\rm s}$ and how
does it change (e.g., a sharp break or a gradual transition)?  In addition,
simultaneous observations at multiple frequencies are important; a dynamical
timescale of our target region is an order of $t_{\rm dyn}\sim 10~R_{\rm s}/c \sim
7$~days, so the jet structure (size and position of the core) can be variable on
this timescale~\citep[e.g.,][]{acciari2009}. To address these issues, we are
currently conducting new high-sensitivity VLBA observations at 86~GHz in
combination with quasi-simultaneous sessions at lower frequencies, which allows
more robust investigations of the jet structure within $100~R_{\rm s}$ and thus
test some specific models more quantitatively. Finally, we also stress the
significance of future imaging opportunities with RadioAstron at 22~GHz or
global-VLBI at 43/86~GHz including ALMA baselines, because these will provide
images at drastically improved resolutions, especially in the direction transverse
to the M87 jet.

\acknowledgments 

We acknowledge the anonymous referee for his/her careful review and suggestions
for improving the paper. We thank K.~Asada and M.~Nakamura for valuable
discussion. We are also grateful to M.~Takahashi, A.~Tomimatsu, M.~Rioja,
R.~Dodson, Y.~Asaki, S.~Mineshige, S.~Kameno, K.~Sekimoto, T.~Tatematsu and
M.~Inoue for useful comments. The Very-Long-Baseline-Array is operated by National
Radio Astronomy Observatory, a facility of the National Science Foundation,
operated under cooperative agreement by Associated Universities, Inc. This work
made use of the Swinburne University of Technology software
correlator~\citep{deller2011}, developed as part of the Australian Major National
Research Facilities Programme and operated under license. This research has made
use of data from the MOJAVE database that is maintained by the MOJAVE
team~\citep{lister2009}. This work was partially supported by KAKENHI (24340042
and 24540240). Part of this work was done with the contribution of the Italian
Ministry of Foreign Affairs and University and Research for the collaboration
project between Italy and Japan. KH is supported by Canon Foundation between April
2012 and March 2013, and by the Research Fellowship from the Japan Society for the
Promotion of Science (JSPS) from April 2013.

\end{document}